\documentclass[superscriptaddress,twocolumn, showpacs,amsmath,amssymb,aps,prl,floatfix, 10pt]{revtex4-2}
\usepackage[american]{babel}
\usepackage[T1]{fontenc}
\usepackage{mathrsfs}
\usepackage{hyperref}
\usepackage{physics}
\usepackage{mathtools}
\usepackage{multirow}
\usepackage{bm, epsfig}
\usepackage{graphicx}
\usepackage{subeqnarray}
\usepackage{bbold}
\usepackage{upgreek}
\usepackage{tikz}
\usepackage{slashed}
\usepackage{lineno}
\usepackage{longtable}
\usepackage{booktabs}
\usepackage{float}

\hypersetup{colorlinks=true, urlcolor=blue, citecolor=blue, filecolor=blue, linkcolor=blue}

\newcommand{\Li}[2]{\mathrm{Li}_{#1}\left(#2\right)}

\newcommand{\rmsZr}{4.2732(7)}
\newcommand{\cZr}{4.9057(77)}
\newcommand{\aZr}{0.5259(35)}
\newcommand{\Zrchi}{0.995 }
\newcommand{\rmsSn}{4.6518(34)}
\newcommand{\Snchi}{0.88 }

\DeclareUnicodeCharacter{202F}{ }

\begin{document}

\title{Relativistic recoil as a key to the fine-structure puzzle in muonic $^{90}\text{Zr}$}

\date{\today}

\author{Konstantin~A.~Beyer}
\affiliation{Max Planck Institute for Nuclear Physics,  Saupfercheckweg 1, 69117 Heidelberg, Germany}
\author{Igor~A.~Valuev}
\affiliation{Max Planck Institute for Nuclear Physics,  Saupfercheckweg 1, 69117 Heidelberg, Germany}
\author{Zoia~A.~Mandrykina}
\affiliation{Max Planck Institute for Nuclear Physics,  Saupfercheckweg 1, 69117 Heidelberg, Germany}
\author{Zewen~Sun}
\affiliation{Max Planck Institute for Nuclear Physics,  Saupfercheckweg 1, 69117 Heidelberg, Germany}
\author{Natalia~S.~Oreshkina}
\email[Email: ]{Natalia.Oreshkina@mpi-hd.mpg.de} 
\affiliation{Max Planck Institute for Nuclear Physics,  Saupfercheckweg 1, 69117 Heidelberg, Germany}

\begin{abstract}
The long-standing fine-structure anomaly in muonic $^{90}\text{Zr}$ is resolved through a rigorous treatment of the relativistic-recoil effect.
From a fit of ab initio QED calculations of the muonic $^{90}$Zr spectrum to precision measurements performed four decades ago, we extract a significantly more precise root-mean-square (rms) charge radius with 6-fold improvement in quality of the fit.
A 2-parameter Fermi (2pF) distribution is assumed to model the nuclear charge density and yields a best-fit value of rms charge radius of $r_\text{rms}[^{90}\text{Zr}]=\rmsZr$~fm ($\chi^2 /{\text{DoF}} = \Zrchi$), in agreement with the previous muonic spectroscopy value, but a factor $6$ more precise, and 3$\sigma$ larger than the accepted literature value. 
Additionally, the same analysis has been performed for {$^{120}$}Sn, where the extracted value of $r_\text{rms}[^{120}\text{Sn}]=\rmsSn$ fm ($\chi^2 /{\text{DoF}} = \Snchi$) is consistent with the accepted value.
These results confirm our assumption that the muonic fine-structure puzzle arose from an incomplete treatment of QED effects and their uncertainties. 

\end{abstract}

\maketitle


{\it Introduction. --- }
Precise nuclear root-mean square (rms) charge radii are a prerequisite for many areas of physics. 
Besides fundamental importance for nuclear, atomic, and molecular physics, they lie at the heart of precision tests of quantum electrodynamics~(QED), the determination of fundamental constants and searches for physics beyond the standard model. 
It is therefore important to extract the rms charge radii accurately and precisely.

Muonic atoms, compared to electronic ones, have an increased sensitivity to nuclear parameters because of the muon's tighter binding; a result of the increased mass~\cite{Wheeler1949,BorieRinker1982}. 
Therefore, the most accurate measurements of rms charge radii generally come from muonic atom spectroscopy. 
QED calculations are fitted to the precision measurements to extract the nuclear parameters in question. 

The rms charge radius extraction of the heavy elements $\mu-^{90}$Zr, $\mu-^{120}$Sn, and $\mu-^{208}$Pb revealed a striking discrepancy between the precision experimental measurements and the QED calculations \cite{Yamazaki1979,  Phan1985, Bergem1988, Piller1990}, manifested in the poor quality of the fit. 
Originally, this ``fine-structure puzzle" was attributed to nuclear polarization (NP), a notoriously difficult to calculate correction to the spectrum; however, state-of-the-art theory confirmed the anomaly \cite{Valuev2022}. 
In our recent work~\cite{Pb_PhysRevLett.135.163002}, we reported on a systematic reevaluation of all relevant QED corrections for the case of $\mu-^{208}$Pb, which resulted in an approximately twenty-fold improvement in the quality of the overall fit and resulted in an rms charge radius in tension with the previously accepted literature values~\cite{Angeli2013Table, Fricke2004}. 

A resolution of this puzzle would remove the ambiguity for unconventional QED, nuclear, or speculative contributions from new physics \cite{BEYER2024138746}, and would provide a more robust baseline for ongoing precision studies across nuclear, atomic, and fundamental physics~\cite{Dickopf2024,Debierre2022,Viatkina2023,Wilzewski2025}.
Furthermore, it motivates further reevaluation of nuclear radii, based both on muonic-atom spectroscopy~\cite{2025_muonic_Cl, antwis2025comparative} and alternative approaches~\cite{Yerokhin_Pb_2025}.

In this work, we consider the two remaining elements for which the muonic fine-structure anomaly has been observed: zirconium and tin. 
While in general the same QED  effects need to be evaluated, their relative importance strongly depends on the nucleus in question. For example, the relativistic recoil (RRecoil) correction originating from the finite nuclear mass is more pronounced in lighter elements: it becomes dominant compared to, {e.g., NP} for intermediate masses and ultimately plays the key role in the current reevaluation.  
With our updated QED theory,  we achieved a statistically significant improvement in the resulting accuracy and in the quality of the fit, in both cases, $\mu-^{90}$Zr, and $\mu-^{120}$Sn. 
Therefore, the muonic fine-structure anomaly can be considered as effectively resolved. 

{\it Theory. --- }
The spectrum of the muonic cascade depends on a number of nuclear parameters, which are treated as free parameters for the fit to measurements. 
For heavy elements, the expansion in $Z\alpha$, with $Z$ being the charge number, converges slowly and thus the Dirac equation must be solved in the full electromagnetic background potential. 
This requires a nuclear charge distribution function $\rho(\vec{r})$, and customarily a 2-parameter Fermi~(2pF) distribution is assumed:
\begin{equation}
	\rho_\mathrm{2pF}(\vec{r}) = \frac{Z}{-8\pi a^3 \mathrm{Li}_3\left(-\exp\left(c/a\right)\right)}\frac{1}{1+\exp\left((r-c)/(a)\right)},
\end{equation}
where $c$ and $a$ are the free parameters, and $\mathrm{Li}_n(x)$ is the polylogarithm function analytically continued to the relevant negative real axis. 
It is important to keep in mind that 2pF is merely a simple, analytic function we use due to a lack of reliable charge distributions predicted from first principles. 

Besides the Coulomb potential, numerous QED corrections affect the spectrum. Some of them can be incorporated at the level of the potential, while others must be treated perturbatively. All the corrections are calculated as an expansion in the fine-structure constant $\alpha$, but simultaneously to all orders in $Z\alpha$.

To leading order in $\alpha$, there are two QED corrections: vacuum polarization (VP) and self-energy (SE). 
The contribution from VP can be expressed as a sum of two contributions: Uehling (Ue) \cite{uehling} and Wichmann-Kroll (WK)~\cite{Wichmann:1956zz} corrections, both of which are included on the level of the potential. 
The dominant Ue corrections are calculated rigorously for a 2pF distribution following~\cite{Klarsfeld1997Analytical}. 
Since numerical precision is the only limiting factor, we ensured that the error is negligible compared to the experimental uncertainty. 
For the evaluation of the WK contribution including the finite-nuclear-size~(FNS) effect, two independent methods have been used, following Refs.~\cite{Salman2023wk,Ivanov2024wk} and Refs.~\cite{Yerokhin_FNS_Lamb_2011, Zaytsev2024qed}. 
Moreover, we have also included the K\"all\'en-Sabry (KS) contribution~\cite{Kallen:1955fb}, which accounts for the second-order correction beyond the Uehling term, following the method described in Ref.~\cite{Indelicato2013Nonperturbative}. 
Even though exponentially suppressed by the loop-particles' masses, muonic-Ue and hadronic-Ue have been included~\cite{Breidenbach_PhysRevA.106.042805,mandrykina_2025hadronicvacuumpolarizationeffect}.

The SE contribution cannot be expressed as a potential term and therefore must be added perturbatively. 
It has been calculated within a fully rigorous QED framework, following  Ref.~\cite{Oreshkina2022}. For the uncertainties, we investigated the dependence of SE on the nuclear charge distribution model. 
A similar investigation but varying Skyrme parametrizations of the nuclear forces has been applied for the nuclear-polarization (NP) correction, which has been calculated in the same manner as described in Ref.~\cite{Valuev2022}, now for all states appearing in the spectra.
The uncertainties have been conservatively estimated as the spread of the results for the nine different Skyrme parametrizations that have been used in Ref.~\cite{Valuev2022} and shown to cover a wide range in the parameter space. 

Finally, the relativistic recoil (RRecoil) correction has to be calculated rigorously. We adopted the approach of Ref.~\cite{Yerokhin2023recoil} and improved it by incorporating a 2pF charge distribution instead of the previously used exponential model, taking the dependence on the model as a proxy for the uncertainty. 
As it holds the key to the solution of the discrepancy between experiment and previous theory calculation, we will go into further detail in the next section.

In Table~\ref{tab:corrections} we present all relevant corrections, calculated perturbatively. 
For the fit to the data, all VP corrections were included non-perturbatively.

\begin{table*}
    \begin{tabular}
{p{1.2cm}p{2.0cm}p{1.2cm}p{1.2cm}p{1.2cm}p{1.2cm}p{1.2cm}p{1.5cm}p{1.6cm}p{1.6cm}p{1.3cm}}
    \hline\hline
    State &	2pF binding\phantom{1} energy & Ue & WK & KS & $\mu$Ue & HadUe & SE & NP & RRecoil & Screening \\
    \hline
    $1s_{1/2}$ & 3645.980 & 25.568 & -0.0526 & 0.2075 & 0.0792 & 0.0533 & -1.239(20) & 1.481(154) & -3.2866(90) & 2.276  \\ 
    $2s_{1/2}$ & 1022.283 & \phantom{1}4.972 & -0.0171 & 0.0383 & 0.0115 & 0.0077 & -0.198(3) & 0.203(21) & -1.0973(17) & 2.274 \\
    $2p_{1/2}$ & 1149.163 & \phantom{1}5.906 & -0.0227 & 0.0432 & 0.0017 & 0.0011 & -0.005(2) & 0.072(10) & -1.4206(11) & 2.275 \\ 
    $2p_{3/2}$ & 1128.390 & \phantom{1}5.577 & -0.0220 & 0.0406 & 0.0011 & 0.0007 & -0.048(1) & 0.066(9) & -1.4037(5) & 2.275 \\ 
    $3p_{1/2}$ & \phantom{1}508.623 & \phantom{1}1.804 & -0.0086 & 0.0133 & 0.0006 & 0.0004 & -0.006(4) & 0.023(3) & -0.6329(3) & 2.268 \\
    $3p_{3/2}$ & \phantom{1}502.672 & \phantom{1}1.722 & -0.0085 & 0.0126 & 0.0004 & 0.0003 & -0.015(1) & 0.021(3) & -0.6285(2) & 2.268 \\
    $3d_{3/2}$ & \phantom{1}503.715 & \phantom{1}1.428 & -0.0088 & 0.0010 & 0.0000 & 0.0000 &  \phantom{-}0.003(2) & 0.0015(2) & -0.6339(1) & 2.270 \\ 
    $3d_{5/2}$ & \phantom{1}501.316 & \phantom{1}1.401 & -0.0087 & 0.0010 & 0.0000 & 0.0000 & -0.003(2) & 0.0015(3) & -0.6309(0) & 2.270 \\ 
    \hline\hline
    \end{tabular} 
    \caption{ Corrections to muonic $^{90}\mathrm{Zr}$  binding energies (keV) calculated perturbatively with FNS wavefunctions from the 2pF model for the best-fit values $a=\aZr$ fm, $c=\cZr$ fm, $\text{rms radius}=\rmsZr$ fm. 
    A positive correction refers to an increase in binding energy. For muonic $^{120}\mathrm{Sn}$, see Appendix.}
    \label{tab:corrections}
\end{table*}

\begin{figure}
    \centering
    \includegraphics[width=0.99\linewidth]{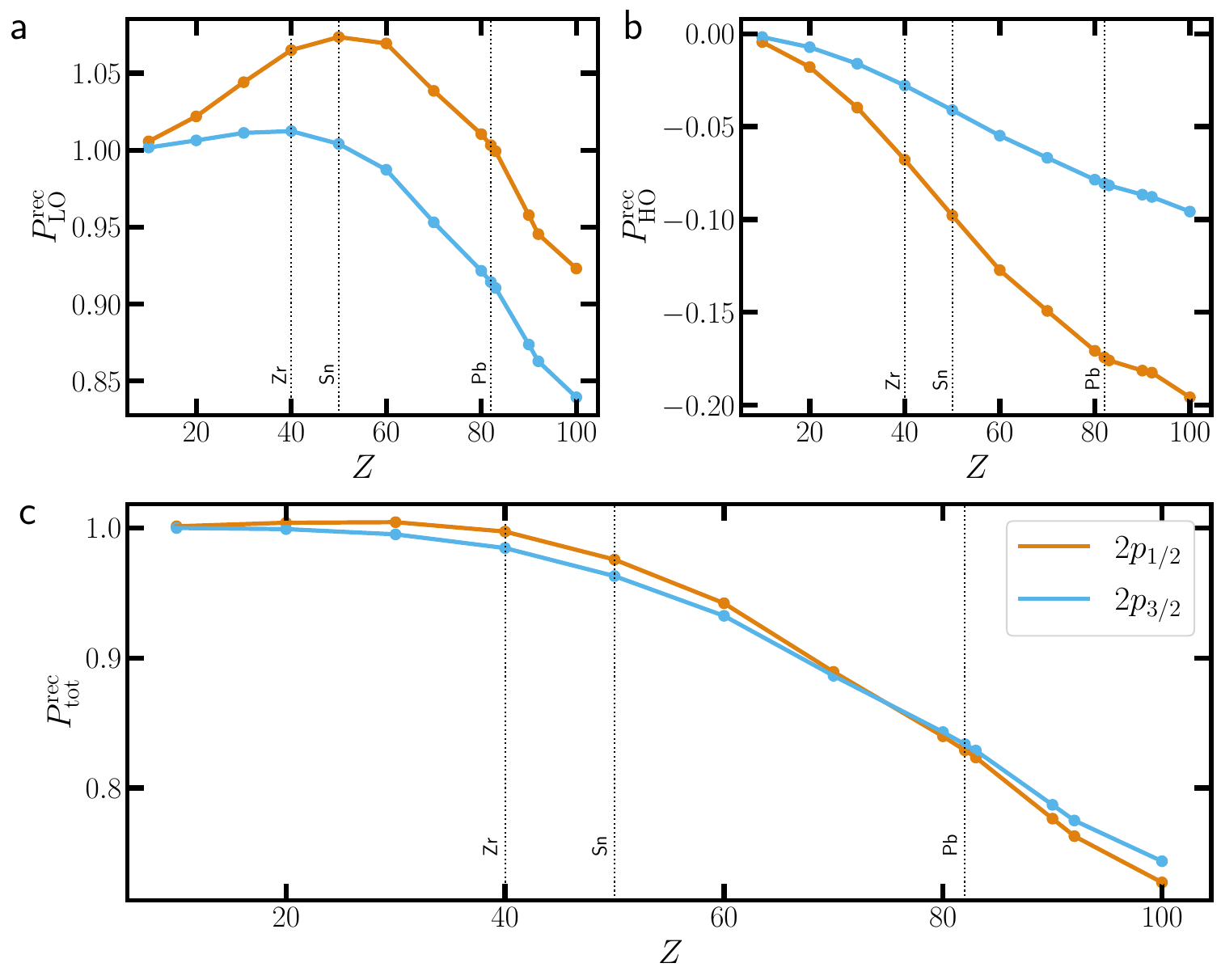}
    \caption{The figure depicts the energy shift from RRecoil as calculated in \cite{Yerokhin2023recoil}. The plots show a) the lower order correction, b) the higher order corrections, and c) the total RRecoil{, see also Eqs.~\eqref{eq:recoil_total}-\eqref{eq:recoil_lo}}. The orange (blue) curve shows the correction for the $2p_{1/2}$ ($2p_{3/2}$) state. It is evident, that the hierarchy for high $Z$ elements, like Pb, is inverted compared to low and moderate $Z$ elements like Zr and Sn. }
    \label{fig:RelRecoil_data}
\end{figure}

{\it Relativistic recoil effect. ---}
The most significant improvement in the current work compared to the previous analysis \cite{Phan1985} is the inclusion of the relativistic recoil effect in a rigorous way. 
Previously, the recoil has been taken to ``all-orders'' in muon-to-nucleus mass ratio $m_\mu/m_{\rm nucl}$ by including the reduced mass in the Dirac equation as a ``non-relativistic term'', while a ``relativistic term'' has been estimated according to the approach described in~\cite{BorieRinker1982}.
Unfortunately, this procedure, inspired by the non-relativistic two-body Schr\"odinger equation, is inconsistent for the relativistic two-body Dirac equation counterpart and therefore mathematically incorrect. 

Furthermore, all-orders in $Z\alpha$ must be taken into account: the bound muons are highly relativistic and the $Z\alpha$ expansion is known to converge slowly even for light systems \cite{Pachucki_PhysRevLett.130.053002}.
Finally, the finite nuclear size has to be taken into account since the Bohr radius of the bound muon is of the same order of magnitude as the nuclear radius for both nuclei under consideration.

To tackle all the difficulties listed above we have used fully-relativistic rigorous QED to calculate the relativistic recoil effect with a finite-size nucleus adapted from~\cite{Yerokhin2023recoil}
\begin{equation}\label{eq:recoil_total}
    E_{\text{RRecoil}} = \frac{i}{2\pi M_{\text{nucl}}}\int_{-\infty}^{\infty} d\omega \sum_n \frac{|\langle a |\vec{p}-\vec{D}(\omega)|n\rangle|^2 }{E_a+\omega -E_n(1-i0)}.
\end{equation}
Here, $M_{\text{nucl}}$ is the nuclear mass, $\ket{a}$ the muonic state with the energy $E_a$, and $\vec{D}$ is obtained from
the transverse part of the photon propagator (see~\cite{Yerokhin2023recoil} for details).

In the same way as in our previous work \cite{Pb_PhysRevLett.135.163002}, we used a 2pF nuclear charge distribution for the leading contribution
\begin{equation}\label{eq:recoil_lo}
    E_{\text{Rrecoil, LO}} = \frac{1}{2M_{\text{nucl}}}\langle a |\vec{p}^2|a\rangle.
\end{equation}

The subleading, or higher-order, contributions (HO), containing the higher-order Coulomb-photon and transverse photon exchange, cannot be calculated for a 2pF nuclear charge distribution with reasonable effort. We therefore use the existing evaluation based on an exponential nuclear charge model of~\cite{Yerokhin2023recoil} and associate an uncertainty to the RRecoil which we define as the relative difference between the LO predictions of the two nuclear charge distributions scaled to the HO contribution.

The result is commonly represented in the form of a dimensionless function $P$ as 
\begin{equation}
    E_{\text{rec}} = m_\mu c^2\frac{m_\mu}{M_{\text{nucl}}}\frac{(\alpha Z)^2}{2n^2} P_{\text{rec}}(\alpha Z).
\end{equation}
As can be seen in Fig.~\ref{fig:RelRecoil_data}{,} the total RRecoil correction shows an inversion of the hierarchy between the corrections to the $2p_{1/2}$ and $2p_{3/2}$ muonic states for Pb compared to Zr and Sn, which is not present in the LO contribution alone, and ultimately results in the resolution of the fine-structure anomaly of $^{90}$Zr.

{\it Experimental data. --- }
The experimental measurements have been performed by \cite{Phan1985} for zirconium and \cite{Piller1990} for tin. The transition energies can be found in Table \ref{tab:exp}. Previously, the data set had been used to fit for the 2pF parameters $c$ and $a$, and a number of NP corrections because of a poor fit using theoretically motivated NP values (see their Table VII). 
We take their measurements and experimental uncertainties as the basis for our fit. 

\begin{table}[h!]
    \centering
    \begin{tabular}{p{2cm} p{3cm} p{2cm}}
    \hline\hline
    Set & Transition &	Energy (keV)  \\
    \hline
    Run I & $2p_{3/2} - 1s_{1/2}$ & 2536.237(22)\\ 
    & $2p_{1/2} - 1s_{1/2}$ & 2515.122(23)\\[1.5mm]
    Run II & $3d_{3/2} - 2p_{1/2}$ & \phantom{1}649.229(10)\\ 
    & $3d_{5/2} - 2p_{3/2}$ & \phantom{1}630.533(9)\\ 
    & $3d_{3/2} - 2p_{3/2}$ & \phantom{1}628.112(12)\\[1.5mm] 
    Run III & $3p_{3/2} - 2s_{1/2}$ & \phantom{1}522.470(24)\\ 
    & $3p_{1/2} - 2s_{1/2}$ & \phantom{1}516.419(17)\\ 
    & $2s_{1/2} - 2p_{1/2}$ & \phantom{1}127.525(12)\\ 
    & $2s_{1/2} - 2p_{3/2}$ & \phantom{1}106.404(9)\\
    \hline\hline
    \end{tabular}
    \caption{Experimentally measured transitions for $^{90}$Zr with their error bars~\cite{Phan1985}.  For muonic $^{120}\mathrm{Sn}$, see Appendix.}
    \label{tab:exp}
\end{table}

{\it Fitting procedure. --- } 
We performed a simple least square error fit to extract the free parameters $c$ and $a$. The square error is defined as
\begin{equation}
\label{Eq:chi2}
    \chi^2(c,a) =\Delta\mathbf{E}^T(c,a)\cdot \sigma^{-1}\cdot\Delta\mathbf{E}(c,a),
\end{equation}
where $\Delta\mathbf{E}$ is the vector of residuals (difference between theory and experiment for all measured, linearly independent transitions), and $\sigma^{-1}$ is the inverse covariance matrix. The latter combines the experimental uncertainties, assumed to be diagonal for the transitions, with the theoretical uncertainties stemming from NP, SE, and RRecoil. 

\begin{figure}
    \centering
    \includegraphics[width=0.9\linewidth]{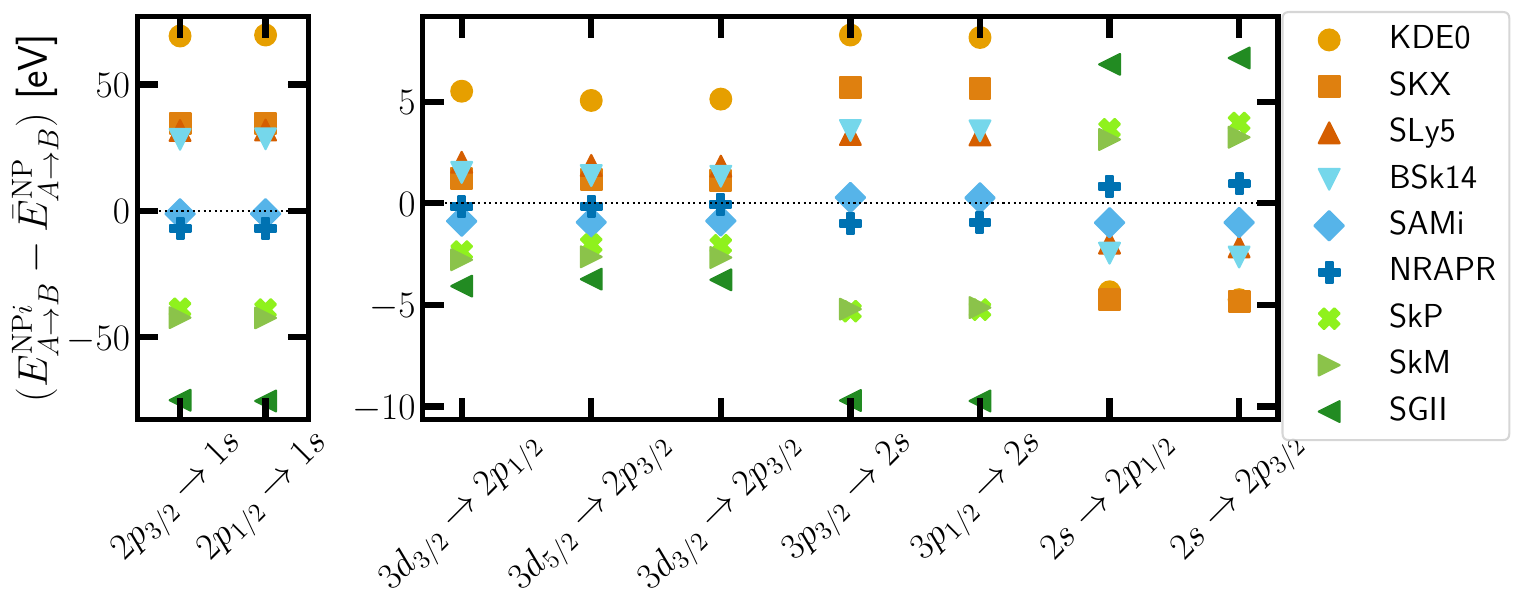}
    \caption{This plot shows the various NP models {for $^{90}$Zr} considered for the ensemble average. Each model predicts a set of NP corrections to the various transitions. All those are plotted against the mean value here. }
    \label{fig:NP_data}
\end{figure}

The covariance matrix for NP can be approximated through an ensemble of NP data for different nuclear models, following \cite{Valuev2022}. A selection of Skyrme-type nuclear potentials was chosen to guarantee a wide spread in parameter space and the average value was used as the central value for the fit (see Fig.~\ref{fig:NP_data}). The covariance matrix is then calculated as the unbiased standard deviation of the ensemble. We inflate the NP covariance by a factor 2 to account for possible bias stemming from both the Skyrme potential treatment and models which were missed.

For the SE correction, we have estimated the level of uncertainty in the theoretical calculation as a combination of the spread between different models, integration techniques and convergence issues.
Because of the nature of these uncertainties and the observation that they are generally small compared to NP uncertainties, we have limited ourselves to an estimation of the physical correlation arising from the fact that multiple transitions test the same underlying states.

We estimated RRecoil covariance by scaling the nuclear model dependence of the LO correction to the HO correction which can not easily be calculated for the 2pF distribution and is calculated for an exponential distribution instead:
\begin{equation}
	\delta E_\mathrm{RRecoil} = \frac{\lvert E_\mathrm{RRecoil,LO}^\mathrm{exp} - E_\mathrm{RRecoil,LO}^\mathrm{2pF}\rvert}{E_\mathrm{RRecoil,LO}^\mathrm{exp}} E_\mathrm{RRecoil,HO}^\mathrm{exp}.
\end{equation}

Minimization of \eqref{Eq:chi2} then results in the best-fit Fermi parameters $c$ and $a$ from which the rms charge radius can be extracted
\begin{equation}
    r_\mathrm{rms}^2(c,a) = 12 a^2\frac{\Li{5}{-\mathrm{exp}\left(c/a\right)}}{\Li{3}{-\mathrm{exp}\left(c/a\right)}}.
\end{equation}

For further detail we refer the reader to our previous publication \cite{Pb_PhysRevLett.135.163002}.

\begin{figure*}
    \centering\includegraphics[width=15cm]{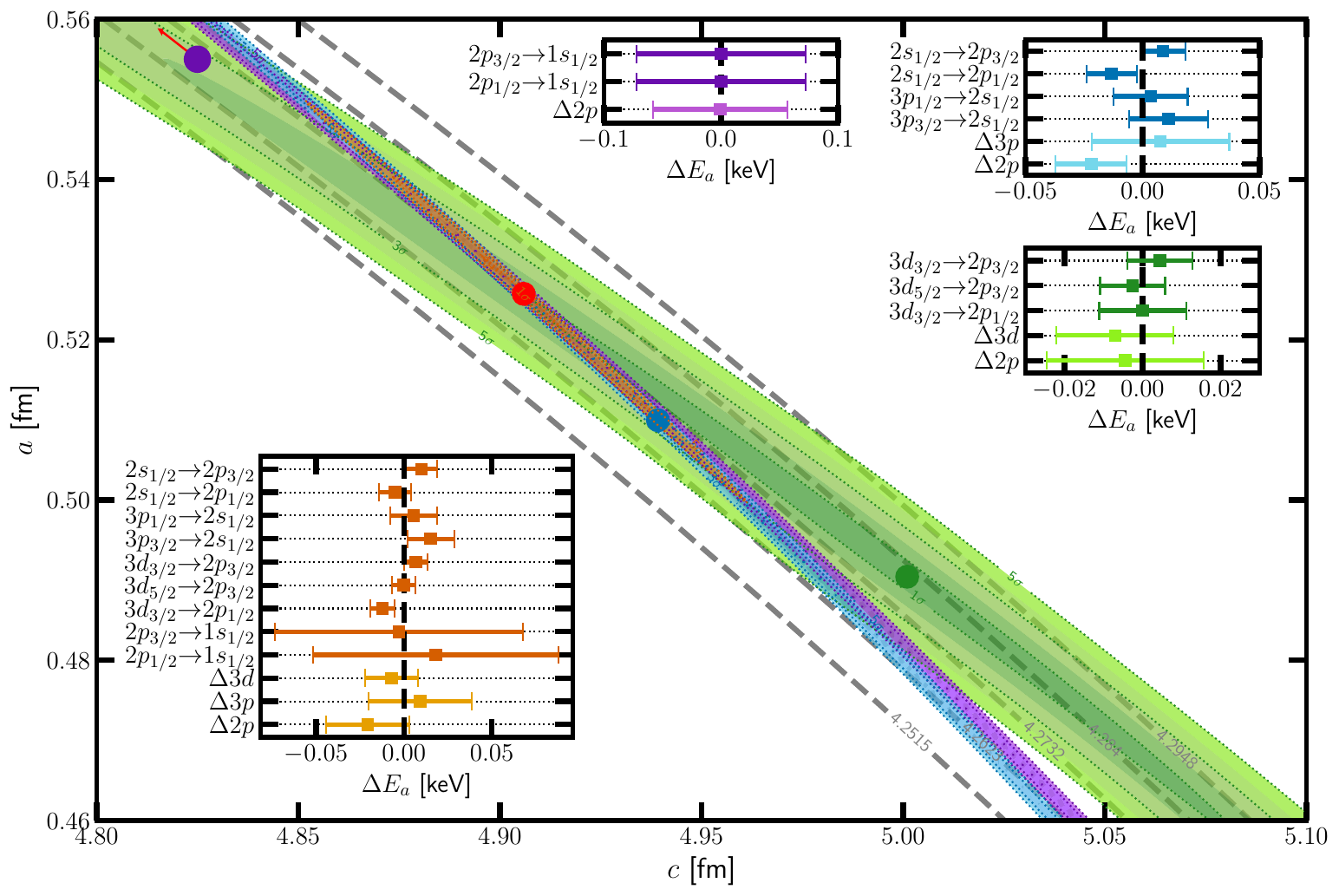}
    \caption{The fit of the spectrum of $\mu-^{90}$Zr to a 2pF distribution. The colored regions show the $5\sigma$ likelihood for the 2pF parameters and the dotted contours highlight the $5$, $3$, and $1\sigma$ steps. Each colour corresponds to a fit of different groups of transitions: purple only fits Run I, green corresponds to Run II, blue to Run III, and orange is the combined fit of all the transitions. The insets show the resulting residual transition energies in the same colours, with measured transitions in darker shade and resulting fine-structure splittings  in lighter colour. The dashed lines correspond to curves of constant radius. The coloured circles indicate the best-fit parameters and the red circle the final values for the overall fit. Note that the best-fit value for Run I lies outside of the plotted parameter range, as indicated by the red arrow.}
    \label{fig:ac_rms}
\end{figure*}

{\it Results and Discussion. --- }
The result of the $\chi^2$-fit for $^{90}$Zr is presented in Fig.~\ref{fig:ac_rms} which shows the best-fit values and likelihood contours for the parameters $c$ and $a$ together with the residual spectra. Even though the individual fits for Run I-III result in different best-fit parameters (see Table~\ref{tab:result}) the likelihood contours overlap on the $1\sigma$ level and the measured spectrum is reproduced well: the quality of the fit is $\chi^2/\text{DoF} = \Zrchi$. The best-fit parameters are $a=\aZr$ fm and $c=\cZr$ fm which correspond to a rms charge radius of $r_\text{rms}=\rmsZr$~fm; in agreement with the previous value from muonic atom spectroscopy~\cite{Phan1985}, but about 3$\sigma$ larger than the commonly accepted one~\cite{Angeli2013Table}.

\begin{table*}[th!]
    \centering
    \begin{tabular}{p{0.9cm}p{1.3cm}p{1.3cm}p{1.3cm}p{1.5cm}p{2.0cm}  p{1.3cm}p{1.3cm}p{1.4cm}p{1.5cm}p{1.5cm}p{1.5cm}}
    \hline\hline
    &&& $^{90}$Zr &&&&& $^{120}$Sn \\
     & Run I & Run II & Run III & Total & Ref.~\cite{Phan1985} & Run I & Run II & Run III & Total & Ref.~\cite{Piller1990} & Ref.~\cite{ENGFER1974509}\\\hline
    $c$ & 4.59(44) & 4.997(67) & 4.947(36) & 4.9057(77) & 4.8791(8) &5.2(1.8)& 5.31(33)& 5.82(19) & 5.51(4) & --- & 5.482(44) \\ 
    $a$ & 0.65(15) & 0.492(25) & 0.506(17) & 0.5259(35) &  0.5367(4) &0.7(7)&0.58(12)& 0.27(22) & 0.496(21) & --- & 0.517(34)\\ 
    rms  & \multirow{ 2}{*}{4.294(28)} & \multirow{ 2}{*}{4.281(8)} & \multirow{ 2}{*}{4.269(4)} & \multirow{ 2}{*}{\rmsZr} & \multirow{ 2}{*}{4.2736(40)} & \multirow{ 2}{*}{4.68(15)}&\multirow{ 2}{*}{4.637(29)}& \multirow{2}{*}{4.61(3)}&\multirow{2}{*}{\rmsSn} & \multirow{2}{*}{4.6522(6)$^1$} & \multirow{2}{*}{4.660(61)$^2$}\\[-4pt]
        radius &&&&&\\
    $^{\chi^2} / _{\mathrm{DoF}}$ & \, --- & 0.21 & 1.18 & \Zrchi &  4 & \, --- &0.56& 1.4 & \Snchi & --- & --- \\ 
    \hline\hline
    \end{tabular}
    \caption{Nuclear parameters of $^{90} \mathrm{Zr}$ and $^{120}$Sn determined from the fits to muonic transition energies, in fm. For value $^1$, quoted error corresponds to the experimental uncertainty. Value $^2$ has been calculated from the reported 2pF parameters with the propagated uncertainty.}
    \label{tab:result}
\end{table*}

The likelihood surfaces reveal the reason for the relatively high precision of our best-fit rms charge radius compared to the Fermi parameters. The level of correlation is very large and while the different runs result in significantly different $c$ and $a$, the values all fall relatively close to a line of constant radius, as indicated by the dashed lines in Fig.~\ref{fig:ac_rms}. The narrow but extended shape of the surfaces also highlights the importance of fitting both parameters $c$ and $a$, as any cut, e.g. along the customary $a=0.5234$ fm, results in a bad overall fit and incorrectly reproduces the constraints coming from the various transition lines.

Through various triangle sums the fine-structure splitting can be extracted. They are shown in Fig.~\ref{fig:ac_rms} in lighter colour to distinguish them from the primary measurements. It can be seen that the previously reported fine-structure anomaly in the $\Delta 2p = E_{2p_{1/2}} - E_{2p_{3/2}}$ splitting no longer is significant. Upon closer inspection it can be seen that the inclusion of the HO RRecoil correction flips the sign of the residual, therefore bringing it in line with the expected hierarchy from experiment. 
Further investigation of the various improvements we made compared to Ref.~\cite{Phan1985} reveals that upon systematic, one-by-one replacement of the newly calculated NP, SE, and RRecoil corrections with the previously reported ones results in good quality fits ($\chi^2/\mathrm{DoF} < \Zrchi$) except for the RRecoil correction. In the latter case the quality of the fit gets worse by almost an order of magnitude. 
We therefore conclude that this correction was at the heart of the anomaly in $\mu-^{90}$Zr. 
It is interesting to highlight that, despite their similarity, the anomalies in Zr and Pb seem to have different origins, as RRecoil is much less significant in $\mu-^{208}$Pb \cite{Pb_PhysRevLett.135.163002}.

For the case of $^{120}$Sn, the experimental data in Ref.~\cite{Piller1990} are  limited.  
Using the two available $2p-1s$ transitions, we obtain a value of $r_{\rm rms}[^{120}{\rm Sn}] = 4.703(23)$~fm, which is noticeably less precise than the earlier muonic-spectroscopy value and the accepted literature values \cite{Angeli2013Table, Fricke2004}. 
A closer examination shows that in the analysis of \cite{Piller1990}, additional transitions (at least $3d-2p$) were included, although no corresponding experimental data were provided.
To resolve this discrepancy, we used even older measurements reported in Ref.~\cite{ENGFER1974509}, and obtained $r_{\rm rms}=4.6518(34)$~fm with $c=5.51(4)$ fm and $a=0.496(21)$~fm (see Appendix for details).
Our result agrees with the accepted literature value, but our uncertainty is nearly twice as large. Given that our analysis benefits from three decades of advances in QED, nuclear theory, and computational methods, this suggests that the earlier uncertainty was likely significantly underestimated.

{\it Conclusions. ---} 
We performed fully rigorous, state-of-the-art QED calculations for the spectrum of muonic zirconium and tin. A fit to the measured spectrum revealed the resolution of the fine-structure anomaly which previously resulted in a poor quality fit of the calculated spectrum to measurements. We argued that the key to the resolution of this discrepancy was the rigorous treatment of the RRecoil correction which had previously been calculated inconsistently.

Based on the good quality fit ($\chi^2/\mathrm{DoF} = \Zrchi$) we extracted the nuclear rms charge radius of $^{90}$Zr, obtaining a value of $r_\text{rms}[^{90}\text{Zr}]=\rmsZr$ fm, which is $3\sigma$ larger than the commonly accepted value.
An equivalent treatment of $^{120}$Sn resulted in $r_\mathrm{rms}[^{120}\text{Sn}] = \rmsSn$ fm ($\chi^2/\mathrm{DoF} = \Snchi$),  consistent with the accepted value.

From our analysis, several key conclusions can be drawn. First, many of the currently accepted rms charge-radius values likely have significantly underestimated uncertainties. Second, contrary to common belief, the nuclear rms radius cannot be reliably extracted from only the $2p-1s$ transitions; higher transitions are at least equally important, even for medium-$Z$ nuclei such as Zr and Sn. And third, new, reliable experimental data are urgently needed.
Moreover, considering  the evolution of {$\chi^2/{\rm DoF}$ from $\approx 1$ for Zr and Sn to 9.7 for Pb, }we can reasonably assume that the remaining tension originates from nuclear effects—most likely related to adopting a more realistic nuclear charge distribution than the conventional 2pF model.

At first glance, the present findings and the previous study on muonic $^{208}$Pb in Ref.~\cite{Pb_PhysRevLett.135.163002} point at different solutions to the fine-structure anomaly, 
which seemingly depends on the exact element under consideration. 
On a deeper level, however, the solution is always the same: rigorous state-of-the-art QED and an honest estimation of the uncertainties.

{\it Acknowledgments. ---} 
This work is part of and funded by the Deutsche Forschungsgemeinschaft (DFG, German Research Foundation) under the Collaborative Research Centre, Project-ID No. 273811115, SFB 1225 ISOQUANT. This article comprises parts of the PhD thesis work of Z. A. M. and Z. S. to be submitted to Heidelberg University.

{\it Author contributions. ---} 
K.A.B. performed the fit and prepared the manuscript.  I.A.V. performed the NP calculations, N.S.O. performed SE calculations, Z.S. and Z.A.M. performed the remaining QED calculations.
N.S.O. initialized and lead the project. All authors have contributed to the discussions and writing.

{\it Competing interests. ---} The authors declare no competing interests.

\bibliography{refs}

\section{Appendix}
\begin{figure}
    \centering
    \includegraphics[width=0.99\linewidth]{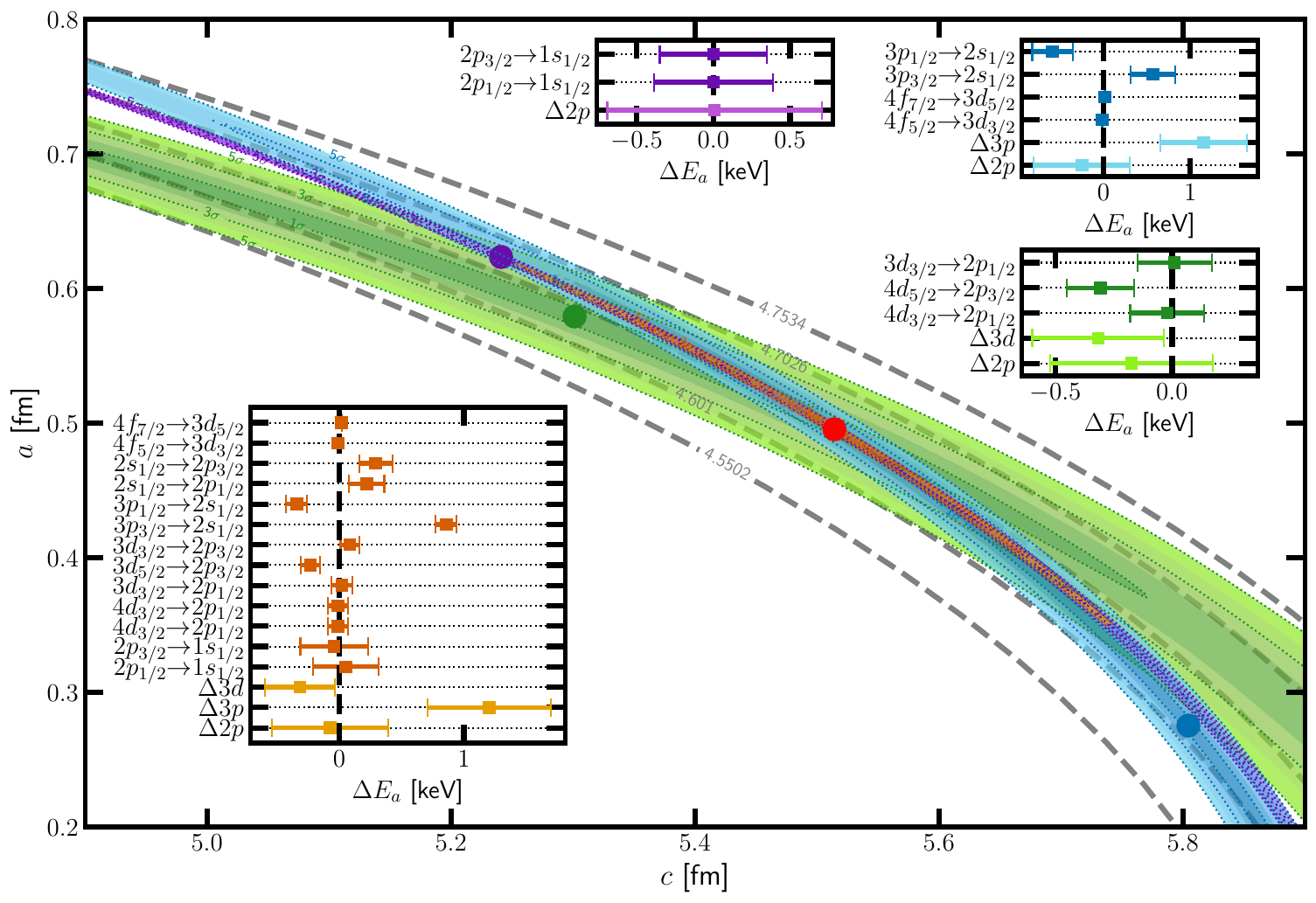}
    \caption{The fit of the spectrum of $\mu-^{120}$Sn to a 2pF distribution. The colored regions show the $5\sigma$ likelihood for the 2pF parameters and the dotted contours highlight the $5$, $3$, and $1\sigma$ steps. Each colour corresponds to a fit of different groups of transitions: purple only fits Run I, green corresponds to Run II, blue to Run III, and orange is the combined fit of all the transitions. The insets show the resulting residual transition energies in the same colours, with measured transitions in darker shade and resulting fine-structure splittings  in lighter colour. The dashed lines correspond to curves of constant radius. The coloured circles indicate the best-fit parameters and the red circle the final values for the overall fit. Note that the best-fit value for Run I lies outside of the plotted parameter range, as indicated by the red arrow.}
    \label{fig:fit_tin}
\end{figure}

Here, we list the experimental data for $^{120}$Sn from Refs.~\cite{Piller1990, ENGFER1974509} in Table~\ref{tab:exp_tin}, provide the complete list of individual contributions for the high-lying states of muonic tin in Table~\ref{tab:sn_corrections}, and show the resulting fit of the spectrum in Fig.~\ref{fig:fit_tin}. {The fitting procedure is analogous to the procedure described in the main body of the text in the context of muonic Zr.}

\begin{table*}
    \begin{tabular}
{p{1.2cm}p{2.0cm}p{1.2cm}p{1.2cm}p{1.2cm}p{1.2cm}p{1.2cm}p{1.5cm}p{1.6cm}p{1.6cm}p{1.3cm}}
    \hline\hline
    State &	2pF binding \phantom{1} energy & Ue & WK & KS & $\mu$Ue & HadUe & SE & NP & RRecoil & Screening \\
    \hline
    $1s_{1/2}$ & 5190.287 & 35.719 & -0.109 & 0.292 & 0.119 & 0.081 & -1.770(40) & 2.547(319) & -3.096(21) & 2.922 \\
    $2s_{1/2}$ & 1527.890 & \phantom{1}7.737 & -0.040 & 0.060 &  0.018 & 0.012 & -0.305(10) &  0.369(46) & -1.157(5) & 2.919 \\
    $2p_{1/2}$ & 1805.319 & 10.437 & -0.056 & 0.078 & 0.005 & 0.004 & -0.036(6) & 0.246(38) & -1.632(1) & 2.920 \\
    $2p_{3/2}$ & 1760.404 & \phantom{1}9.719 & -0.054 & 0.072 & 0.004 & 0.002 & -0.120(5) & 0.227(34) & -1.615(1) & 2.920 \\   
    $3p_{1/2}$ & \phantom{1}798.308 &  \phantom{1}3.266 & -0.022 & 0.025 & 0.002 & 0.001 & -0.025(5) & 0.079(16) & -0.731 & 2.912 \\
    $3p_{3/2}$ & \phantom{1}785.745 & \phantom{1}3.093 & -0.022 &  0.022 & 0.001 & 0.001 & -0.041(2) & 0.076(14) & -0.728 & 2.912 \\
    $3d_{3/2}$ & \phantom{1}790.229 & \phantom{1}2.681 &  -0.024 &  0.019 & 0.000 &  0.000 & \phantom{-}0.005(3) & 0.007(2) & -0.744 & 2.915\\ 
    $3d_{5/2}$ & \phantom{1}784.344 & \phantom{1}2.607 & -0.023 &  0.018 &  0.000 & 0.000 & -0.008(2) & 0.006(3) & -0.739 & 2.915 \\ 
    $4f_{5/2}$ & \phantom{1}441.717 & \phantom{1}0.890 &  -0.011 & 0.006 & 0.000 & 0.000 & \phantom{-}0.000(1) & 0.000(0) & -0.423 & 2.904 \\ 
    $4f_{7/2}$ & \phantom{1}440.485 & \phantom{1}0.879 & -0.011 & 0.006 & 0.000& 0.000 & \phantom{-}0.000(2) & 0.000(0) & -0.420 & 2.904 \\ 
    \hline\hline
    \end{tabular} 
    \caption{Corrections to muonic $^{120}\mathrm{Sn}$ binding energies (keV) calculated perturbatively with FNS wavefunctions from the 2pF model for the best-fit values     $a=0.496(21)$~fm, $c=5.51(4)$~fm, $r_{\rm rms}=\rmsSn$~fm.
    A positive correction refers to an increase in binding energy. }
    \label{tab:sn_corrections}
\end{table*}

\begin{table}[ht!]
    \centering
    \begin{tabular}{p{2cm} p{3cm} p{2cm}}
    \hline\hline
    Set & Transition &	Energy (keV)  \\
    \hline
    \cite{Piller1990} & $2p_{3/2} - 1s_{1/2}$ & 3454.453(50) \\
    & $2p_{1/2} - 1s_{1/2}$ & 3408.975(52) \\[1.5mm]
    Run I \cite{Emma2010} & $2p_{3/2} - 1s_{1/2}$ & 3454.41(33)\\ 
    & $2p_{1/2} - 1s_{1/2}$ & 3408.79(37)\\[1.5mm]
    Run II \cite{Emma2010} & $4d_{3/2} - 2p_{1/2}$ & 1369.50(26)\\ 
    & $4d_{5/2} - 2p_{3/2}$ & 1325.9(5) \\ 
    & $3d_{3/2} - 2p_{1/2}$ & 1022.20(20)\\ 
    & $3d_{5/2} - 2p_{3/2}$ & \phantom{1}982.20(20)\\ 
    & $3d_{3/2} - 2p_{3/2}$ & \phantom{1}976.55(30)\\[1.5mm] 
    Run III \cite{Emma2010} & $3p_{3/2} - 2s_{1/2}$ & \phantom{1}747.15(35)\\ 
    & $3p_{1/2} - 2s_{1/2}$ & \phantom{1}733.20(35) \\    
    & $2s_{1/2} - 2p_{1/2}$ & \phantom{1}280.14(45) \\ 
    & $2s_{1/2} - 2p_{3/2}$ & \phantom{1}234.50(30) \\
    & $4f_{5/2} - 3d_{3/2}$ & \phantom{1}350.0(3) \\
    & $4f_{7/2} - 3d_{5/2}$ & \phantom{1}345.3(3) \\
    \hline\hline
    \end{tabular}
    \caption{Experimentally measured transitions with their error bars~\cite{Piller1990,ENGFER1974509}. }
    \label{tab:exp_tin}
\end{table}

As described in the main text, the initial analysis has been performed with more recent experimental data \cite{Piller1990}; however, because only 2 transitions were measured, the resulting rms charge radius can not be ascertained with a great accuracy, see Table~\ref{tab:result} and Fig.~\ref{fig:fit_tin}. 
This problem had already been noted in the earlier works, and, thus, older data had been combined with the new measurements to increase the accuracy.
For reasons of consistency we repeated the analysis with the full set of older experimental data from \cite{ENGFER1974509} to obtain the final rms charge radius reported in the main text.

As one can see in Fig.~\ref{fig:fit_tin}, relatively large uncertainties in $c$ and $a$ do not simply propagate to the rms value. The two parameters are correlated significantly, indicated by the likelihood contours being almost perfectly aligned with the constant rms lines. 
Nevertheless, the uncertainty of our rms charge radius for tin is a factor of 5.5 larger than the previously reported one in \cite{Piller1990} because previously only experimental uncertainties have been included into the evaluation.  
Furthermore, Ref.~\cite{ENGFER1974509} reports only on the 2pF parameters; therefore, our evaluation of the corresponding rms value probably overestimates the uncertainty since the correlation between $c$ and $a$ is not given and therefore neglected.

\end{document}